\documentclass[apj]{emulateapj}

\usepackage{enumerate}
\usepackage{natbib}
\usepackage{graphicx} 
\usepackage{apjfonts}
\usepackage{color}
\slugcomment{Accepted for publication in The Astrophysical Journal Letters}

\shorttitle{Location of $\gamma$-ray Flare Emission in the Jet of \object{OJ287} $>14$\,pc from the Central Engine}
\shortauthors{Agudo et al.}

\begin{document}
\title{Location of $\gamma$-ray Flare Emission in the Jet of the BL~Lacertae Object \object{OJ287} more than $14$\,pc from the Central Engine}

\author{Iv\'{a}n Agudo\altaffilmark{1}, 
            Svetlana G.~Jorstad\altaffilmark{1,2}, 
            Alan P.~Marscher\altaffilmark{1},
            Valeri M.~Larionov\altaffilmark{2,3},
            Jos\'{e} L.~G\'omez\altaffilmark{4}, 
            Anne L\"{a}hteenm\"{a}ki\altaffilmark{5},
            Mark Gurwell\altaffilmark{6}, 
            Paul S. Smith\altaffilmark{7},
            Helmut Wiesemeyer\altaffilmark{8},
            Clemens Thum\altaffilmark{9}, 
            Jochen Heidt\altaffilmark{10}, 
            Dmitriy A.~Blinov\altaffilmark{2,3},
            Francesca D.~D'Arcangelo\altaffilmark{1,11},
            Vladimir A.~Hagen-Thorn\altaffilmark{2,3}, 
            Daria A.~Morozova\altaffilmark{2},
            Elina Nieppola\altaffilmark{5,12},
            Mar Roca--Sogorb\altaffilmark{4}, 
            Gary D.~Schmidt\altaffilmark{13},
            Brian Taylor\altaffilmark{1,14},
            Merja Tornikoski\altaffilmark{5},
            Ivan S.~Troitsky\altaffilmark{2}
           }


\altaffiltext{1}{Institute for Astrophysical Research, Boston University, 
                 725 Commonwealth Avenue, Boston, MA 02215, USA; \email{iagudo@bu.edu}}

\altaffiltext{2}{Astronomical Institute, St. Petersburg State University, Universitetskij Pr. 28, 
                 Petrodvorets, 198504 St. Petersburg, Russia}

\altaffiltext{3}{Isaac Newton Institute of Chile, St. Petersburg Branch, St. Petersburg, Russia}

\altaffiltext{4}{Instituto de Astrof\'{i}sica de Andaluc\'{i}a, CSIC, Apartado 3004, 18080, Granada, Spain}

\altaffiltext{5}{Aalto University Mets\"{a}hovi Radio Observatory, Mets\"{a}hovintie 114, FIN-02540 Kylm\"{a}l\"{a}, Finland}

\altaffiltext{6}{Harvard--Smithsonian Center for Astrophysics, 60 Garden St., Cambridge, MA 02138, USA}

\altaffiltext{7}{Steward Observatory, University of Arizona, Tucson, AZ 85721-0065, USA}

\altaffiltext{8}{Instituto de Radio Astronom\'{i}a Milim\'{e}trica, Avenida Divina Pastora, 7, Local 20,
                  E--18012 Granada, Spain}

\altaffiltext{9}{Institut de Radio Astronomie Millim\'{e}trique, 300 Rue de la Piscine, 
                  38406 St. Martin d'H\`{e}res, France}

\altaffiltext{10}{ZAH, Landessternwarte Heidelberg, K\"{o}nigstuhl, 69117 Heidelberg, Germany}

\altaffiltext{11}{MIT Lincoln Laboratory, 244 Wood Street, Lexington, MA 02421, USA}

\altaffiltext{12}{Finnish Centre for Astronomy with ESO (FINCA), University of Turku, V\"{a}is\"{a}l\"{a}ntie 20, FI-21500 Piikki\"{o}, Finland}

\altaffiltext{13}{National Science Foundation, 4201 Wilson Blvd., Arlington, VA 22230, USA}

\altaffiltext{14}{Lowell Observatory, Flagstaff, AZ 86001, USA}
\begin{abstract}
We combine time-dependent multi-waveband flux and linear polarization observations with sub-milliarcsecond-scale polarimetric images at $\lambda=7$\,mm of the BL Lacertae-type blazar OJ287 to locate the $\gamma$-ray emission in prominent flares in the jet of the source $>14$\,pc from the central engine. 
We demonstrate a highly significant correlation between the strongest $\gamma$-ray and millimeter-wave flares through Monte-Carlo simulations. 
The two reported $\gamma$-ray peaks occurred near the beginning of two major mm-wave outbursts, each of which is associated with a linear polarization maximum at millimeter wavelengths.
Our Very Long Baseline Array observations indicate that the two mm-wave flares originated in the second of two features in the jet that are separated by $>14$\,pc. 
The simultaneity of the peak of the higher-amplitude $\gamma$-ray flare and the maximum in polarization of the second jet feature implies that the $\gamma$-ray and mm-wave flares are co-spatial and occur $>14$\,pc from the central engine.
We also associate two optical flares, accompanied by sharp polarization peaks, with the two $\gamma$-ray events.
The multi-waveband behavior is most easily explained if the $\gamma$-rays arise from synchrotron self-Compton scattering of optical photons from the flares.
We propose that flares are triggered by interaction of moving plasma blobs with a standing shock. 
The $\gamma$-ray and optical emission is quenched by inverse Compton losses as synchrotron photons from the newly shocked plasma cross the emission region. 
The mm-wave polarization is high at the onset of a flare, but decreases as the electrons emitting at these wavelengths penetrate less polarized regions.
\end{abstract}


\section{Introduction}
 The information that $\gamma$-ray observations can provide on the physical properties of active galactic nuclei in general, and blazars in particular, depends on where such $\gamma$-ray emission originates, which is still under debate \citep[e.g.,][]{Marscher:2010p12402}.
Variability studies of $\gamma$-ray and lower frequency emission from blazars provide important insights into this problem by relating time scales of variability with physical sizes in the source.
However, although $\gamma$-ray variability in blazars can occur on very short time-scales \citep[of a few hours, e.g.,][]{Mattox:1997p12917, Foschini:2010p12452}, this does not necessarily imply that high-energy flares take place at short distances ($\ll 1$~pc) from the central engine \citep[see][]{Marscher:2010p12402}. 
In fact, \citet{Lahteenmaki:2003p5657} and \citet{Jorstad:2001p5655, Jorstad:2001b} provided evidence that $\gamma$-ray outbursts may arise in the mm-wave emitting regions $\gtrsim 1$\,pc downstream of the central engine. 
The Large Area Telescope (LAT) onboard the {\it Fermi} Gamma-ray Space Telescope has sufficient sensitivity to test this by providing well-sampled light curves of blazars that allow detailed studies of the timing of $\gamma$-ray flares relative to those at other spectral ranges \citep[e.g.,][]{Abdo:2010p11811, Jorstad:2010p11830,Marscher:2010p11374}.

The technique developed by \citet{Marscher:2010p11374} and \citet{Jorstad:2010p11830} uses ultra-high angular-resolution ($~\sim0.15$\,milliarsecond) monitoring with very long baseline interferometry (VLBI) to resolve the innermost jet regions and monitor changes in jet structure. 
Observations with roughly monthly observations, supplemented by more concentrated campaigns, provide time sequences of total and polarized intensity images of the parsec-scale jet that can be related to variations of the flux and polarization at higher frequencies. 

In this paper, we employ this technique to investigate the location of the flaring $\gamma$-ray emission in the BL~Lacertae (BL~Lac) object \object{OJ287} ($z=0.306$),  a well studied and highly variable blazar at all available spectral ranges \citep[e.g.,][]{Abdo:2010p11947, Abdo:2010p11698}.
Throughout this paper we adopt the standard $\Lambda$CDM cosmology with $H_0$=71 km s$^{-1}$ Mpc$^{-1}$, $\Omega_M=0.27$, and $\Omega_\Lambda=0.73$, so that 1~milliarcsecond (mas) corresponds to a projected distance of 4.48~pc, and a proper motion of 1\,mas/yr corresponds to a superluminal speed of $19\,c$.

\section{Observations}
\begin{figure}
   \centering
   \includegraphics[clip,width=7.cm]{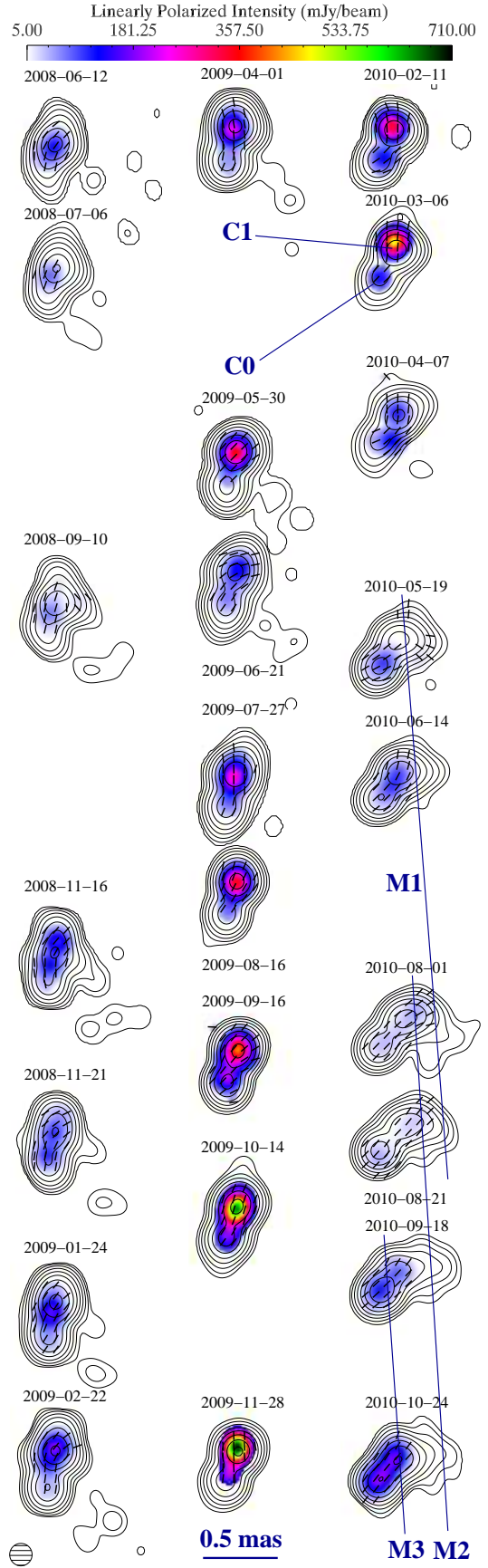}
   \caption{Sequence of 7\,mm VLBA images of OJ287 in 2008--2010.
   Images are convolved with a $\rm{FWHM}=0.15$\,mas circular Gaussian beam.
   Contour levels represent 0.2, 0.4, 0.8, 1.6, 3.2, 6.4, 12.8, 25.6, 51.2, 90.0\% of the peak total intensity of 6.32\,Jy/beam. 
   The color scale indicates linearly polarized intensity, whereas superimposed line segments represent the orientation of the polarization electric-vector position angle.}
   \label{maps}
\end{figure}

Our polarimetric observations of \object{OJ287} include (1) 7\,mm (43\,GHz) Very Long Baseline Array (VLBA) images (Fig.~\ref{maps}), mostly from the Boston University blazar monitoring program\footnote{\tt http://web.bu.edu/blazars/VLBAproject.html}, (2) 3\,mm (86\,GHz) monitoring with the IRAM 30\,m Telescope, and (3) optical ($R$ and $V$ band) photo-polarimetric observations from several observatories (Figs.~\ref{tflux} and \ref{pol}).  
The optical facilities include Calar Alto (2.2\,m telescope, observations under the MAPCAT\footnote{\tt http://www.iaa.es/$\sim$iagudo/research/MAPCAT} program), Steward (2.3 and 1.54\,m telescopes\footnote{Data listing: {\tt http://james.as.arizona.edu/$\sim$psmith/Fermi}}), Lowell (1.83\,m Perkins Telescope), St. Petersburg State University (0.4\,m telescope), and Crimean Astrophysical  (0.7\,m telescope) observatories. 
To these we add $R$-band polarimetric (and $V$-band photometric) data from \citet{Villforth:2010p11557}.
The total flux light curves analyzed here (see Fig.~\ref{tflux}) are from the \emph{Fermi}-LAT $\gamma$-ray (0.1--200\,GeV) and \emph{Swift} X-ray (0.3--10\,keV) and optical ($V$-band) data available from the archives of these missions, by the Yale University SMARTS program\footnote{Data listing: {\tt http://www.astro.yale.edu/smarts/glast}}, by the Submillimeter Array (SMA) at 1.3\,mm (230\,GHz) and 850\,$\mu$m (350\,GHz), by the IRAM 30\,m Telescope at 1.3\,mm, and by the Mets\"{a}hovi Radio Observatory 14\,m Telescope at 8\,mm (37\,GHz).

Our data analysis follows the procedures from previous studies: (1) VLBA: \citet{Jorstad:2005p264}; (2) optical polarimetric data: \citet{Jorstad:2010p11830}; (3) IRAM data: \citet{Agudo:2006p203,Agudo:2010p12104}; (4) SMA: \citet{Gurwell:2007p12057}; (5) Mets\"{a}hovi: \citet{1998A&AS_132_305T}; (6) \emph{Swift}: \citet{Jorstad:2010p11830}; and (7) \emph{Fermi} LAT: \citet{Marscher:2010p11374}. 
For the LAT data, we considered a $15^{\circ}$ radius centered on \object{OJ287} and used the maximum-likelihood routine GTLIKE to model the 0.1--200~GeV spectra from this object, \object{CRATES~J0856$+$2057}, \object{OJ~290}, and \object{1FGL~J0902.4$+$2050} as single power laws. 
We measured the flux in 7-day bins and fixed the slope of the photon spectrum of \object{OJ287} at $-2.4$, \citep[as determined over the first 11 months of LAT observations by][]{Abdo:2010p12082}.

\section{Analysis of Observational Results}

\subsection{7\,mm Jet Structure and Kinematics}
Fig.~\ref{maps} reveals a radically different sub-milliarcsecond-scale jet structure at 7\,mm than found previously in \object{OJ287}. 
\citet{Jorstad:2005p264} reported a bright core with a one-sided jet at a position angle $PA \approx-110^{\circ}$ in 1998--2001.
Our new images show the innermost jet region oriented with $PA$ from $\sim-10^{\circ}$ to $\sim-40^{\circ}$ between 2008 and 2010, among the most extreme shifts in apparent jet direction yet observed \citep[see e.g.,][]{Agudo:2007p132}.
The change in orientation conflicts with the 12-yr period precession model presented by \citet{Tateyama:2004p11950} that predicted the jet to be at $PA \approx-95^{\circ}$ in 2009. 
A more detailed study of this issue and of the origin of jet wobbling in \object{OJ287} will be presented elsewhere.

\citet{Jorstad:2005p264} also reported a set of quasi-stationary features at average locations $\sim0.1$, $\sim0.3$, and $\sim1$\,mas from the innermost jet feature in \object{OJ287}.
Fast superluminal features propagating downstream with speeds $\gtrsim10\,c$ were also observed to move across these quasi-stationary features.
We model the brightness distribution of the source at 7\,mm with a small number of circular Gaussian components.
Our model fits include a bright quasi-stationary feature (C1) $\sim0.2$\,mas from the innermost jet region (C0).
The identification of C0 as the innermost jet feature is justified by the decreasing intensity westward of C1, and by the detection after 2010 March of superluminal motion of features M1 and M2 toward the west-southwest of C1 with speeds of $10.8c\pm1.3c$ and $6.7c\pm1.4c$) and possibly M3, which crossed C1 in 2010 Oct. (preliminary speed of $\gtrsim10\,c$).

\subsection{Flares in the C1 Jet Region at 1\,mm and 7\,mm}
\begin{figure}
   \centering
   \includegraphics[clip,width=8.cm]{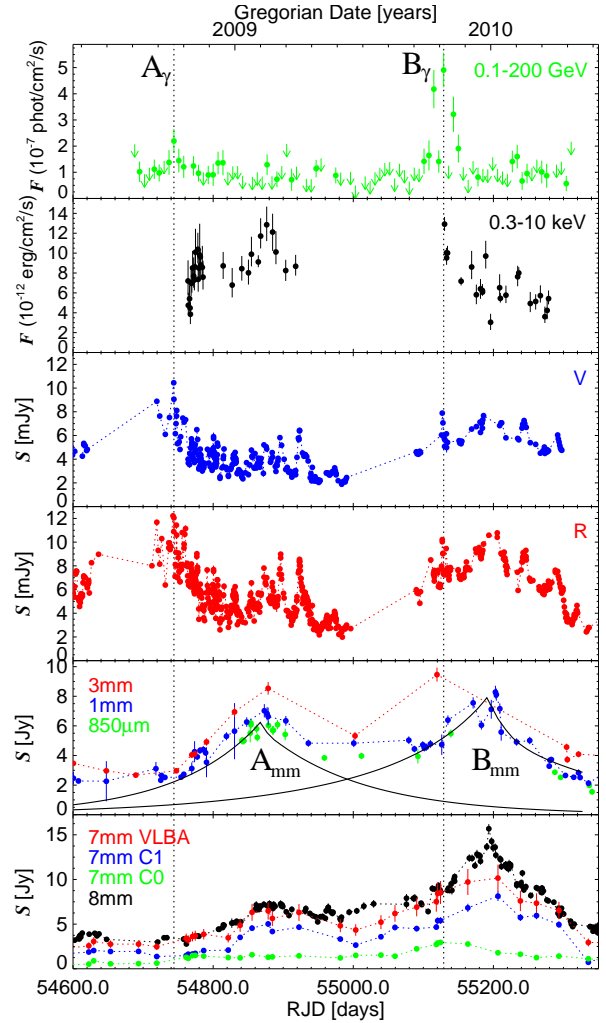}
   \caption{Light curves of OJ287 from mm-wave to $\gamma$-ray frequencies. The vertical lines denote the times of peak $\gamma$-ray flux of ${\rm{A}}_{\gamma}$ and ${\rm{B}}_{\gamma}$.  The solid lines in the next-to-last panel represent fits to the two major 1\,mm flares by using the method of \citet{2008ApJ_689_79C}. RJD = Julian Date $- 2400000.0$.}
   \label{tflux}
\end{figure}

Before M1, M2, and M3 appeared in the jet, C0 and C1 typically contained $\gtrsim90$\% of the 7\,mm flux, and therefore governed the mm-wave evolution, of \object{OJ287}.
In particular, the two most prominent 1\,mm flares ever reported in \object{OJ287} (${\rm{A}}_{\rm{mm}}$ and ${\rm{B}}_{\rm{mm}}$ as labeled in Fig.~\ref{tflux}) took place in C1, as indicated by the correspondence of events in the 7\,mm light curve of this component with those at other millimeter wavelengths.
The multiple peaks of flare ${\rm{B}}_{\rm{mm}}$ may be related to the passage of M1, M2, and M3 through C1, although we cannot verify this given the uncertainties in the trajectories of these knots.
No moving features related to ${\rm{A}}_{\rm{mm}}$ are apparent in the images, although the polarization westward of C1 in the 2009-02-22 and 2009-06-21 images suggests that weak knots were present at those epochs.

From the angle of the jet axis to the line of sight in \object{OJ287} \citep[$1^\circ\kern-.35em .9$--$4^\circ\kern-.35em .1$; see][]{Jorstad:2005p264,Pushkarev:2009p9412}, and the mean projected separation of C1 from C0 at the time of start of ${\rm{A}}_{\rm{mm}}$ and ${\rm{B}}_{\rm{mm}}$ ($0.23\pm0.01$\,mas), we estimate that C1 is located $>14$\,pc downstream of C0, the innermost jet region detected in our images.
The actual distance between the central engine in \object{OJ287} and C1 must be even greater if C0 lies downstream of the acceleration and collimation zone of the jet \citep[ACZ;][]{2007AJ.134.799J,2008Natur_452_966M,Marscher:2010p11374}.

\subsection{$\gamma$-Ray Flares}
The two most pronounced $\gamma$-ray flares take place during the initial rising phases of ${\rm{A}}_{\rm{mm}}$ and ${\rm{B}}_{\rm{mm}}$. 
This conforms with the pattern found in the 1990s by \citet{Lahteenmaki:2003p5657}.
The two $\gamma$-ray flares, peaking on 2008 October 4 (${\rm{A}}_{\gamma}$) and 2009 October 24  (${\rm{B}}_{\gamma}$), reached $0.1$--$200$\,GeV photon fluxes a factor of $\sim2$ and $\sim5$, respectively, higher than the quiescent $\gamma$-ray level ($\sim1\times10^{-7}$ phot/cm$^2$/s). 

The discrete correlation function \citep[DCF;][]{Edelson:1988p12129} between the $\gamma$-ray and 1\,mm long-term light curve (Fig.~\ref{dcf}) possesses a prominent peak at a time lag $\sim -80$ days ($\gamma$-ray leading).
We compute the significance of this result relative to stochastic variability through Monte Carlo simulations as in \citet{2008ApJ_689_79C} and \cite{MaxMoerbeck:2010p12574}.
Our method follows \citet{Timmer:1995p12152} for the simulation of $N=5000$ different pairs of $\gamma$-ray and 1\,mm light curves characterized by the mean and standard deviation of the observed light curves, and by power-law shaped power spectral densities (${\rm{PSD}}\propto1/{f^{a}}$) with slopes in a grid $a_{\gamma}=\{1.0,1.5,2.0\}$ and $a_{\rm{1mm}}=\{1.0,1.5,2.0,2.5,3.0\}$. 
The range of $a_{\gamma}$ values includes the means found by \citet{Abdo:2010p11698}, $a_{\gamma}=1.5$ and $1.7$ for quasars and BL~Lacs, respectively. For $a_{\rm{1mm}}$, we choose values covering the range derived by \citet{2008ApJ_689_79C} and \citet{Hufnagel:1992p13974}.

The results of our simulations show that the DCF peak at a lag of $\sim-80$ days is significant at 99.7\,\% confidence in all of our simulations. 
{\it This confirms the correlation between ${\rm{B}}_{\gamma}$ and ${\rm{B}}_{\rm{mm}}$, the most luminous $\gamma$-ray and 1\,mm flares in our data}.
The correlation of the $\gamma$-ray and 8\,mm light curves in Fig.~\ref{tflux} is of similarly high significance.

The optical light curves (especially $V$-band; Fig.~\ref{tflux}) show two sharp flux increases at essentially zero time lag from ${\rm{A}}_{\gamma}$ and ${\rm{B}}_{\gamma}$.
In contrast, the sparser time coverage of the $0.3$--$10$\,keV light curve does not allow us to make an unambiguous connection between the $\gamma$-ray and X-ray flares.

\subsection{Variability of Linear Polarization}
\begin{figure}
   \centering
   \includegraphics[clip,width=8.cm]{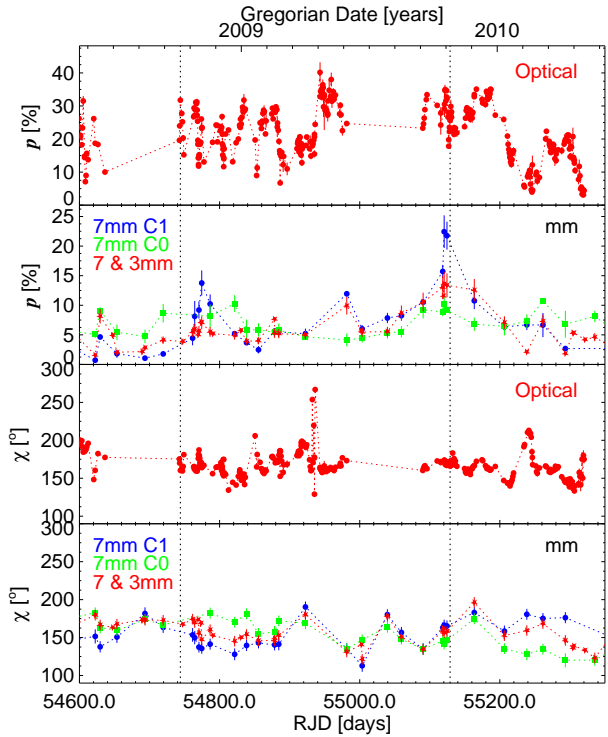}
   \caption{Optical and mm-wave linear polarization of OJ287 as a function of time. The optical data includes R-band and 5000--7000\,{\AA} observations.  The similarity of the integrated linear polarization at 7\,mm and 3\,mm also allows us to combine them. Vertical lines are as in Fig.~\ref{tflux}.}
   \label{pol}
\end{figure}

C0 and C1 dominate the evolution of the linear polarization $p$ and electric vector position angle $\chi$ at 7\,mm in \object{OJ287} (Fig.~\ref{pol}).  
However, whereas $p_{\rm{C0}}$ never exceeds $10$\%, C1 exhibits the two largest peaks in $p$ ever observed in \object{OJ287} at 7\,mm, $p_{\rm{C1}}\approx14$\% on 2008 November 4, and $p_{\rm{C1}}\approx22$\% on 2009 October 16.
The first maximum in $p_{\rm{C1}}$ follows the peak of ${\rm{A}}_{\gamma}$ by one month, while the second more pronounced polarization event is already in progress when the $\gamma$-ray flux of flare ${\rm{B}}_{\gamma}$ rises to a level of $\sim 4\times 10^{-7}$~phot~cm$^{-2}~s^{-1}$. 
{\emph{This coincidence of the strongest $\gamma$-ray outburst and exceptionally strong polarization in C1 identifies this feature $>14$\,pc from the central engine as the site of the variable $\gamma$-ray emission.}}

The optical polarization peaks at essentially the same time as $p_{\rm{C1}}$ at 7\,mm during flare ${\rm{B}}_{\gamma}$. 
During both ${\rm{A}}_{\gamma}$ and ${\rm{B}}_{\gamma}$, $p_{\rm{opt}} \approx 35\%$, which requires a well-ordered magnetic field.
However, comparable optical polarization levels also occur at other times.
The shorter time scale and larger amplitude of variability of optical polarization, as compared with those at millimeter wavelengths, is consistent with frequency dependence in the turbulence model of \citet{Marscher:2010p12402}.

The optical and mm-wave linear polarization position angle is stable at $\chi\approx160^{\circ}$--$170^{\circ}$ ($-20^\circ$ to $-10^\circ$)---similar to the structural position angle of the inner jet---both near ${\rm{A}}_{\gamma}$ and ${\rm{B}}_{\gamma}$ and throughout most of the monitoring period. 
The corresponding direction of the magnetic field is transverse to the direction between features C0 and C1. 
The stability of field direction is only greatly altered in the optical by sporadic short-term rotations of $\chi$ by up to $180^{\circ}$ when the polarization is relatively low \cite[see][]{Villforth:2010p11557}, consistent with the behavior expected if the magnetic field becomes turbulent \citep{Jones:1988p7083,2007ApJ_659L_107D}.

\subsection{Low Probability of Chance Coincidences}
\label{prob}
\begin{figure}
   \centering
   \includegraphics[clip,width=8.cm]{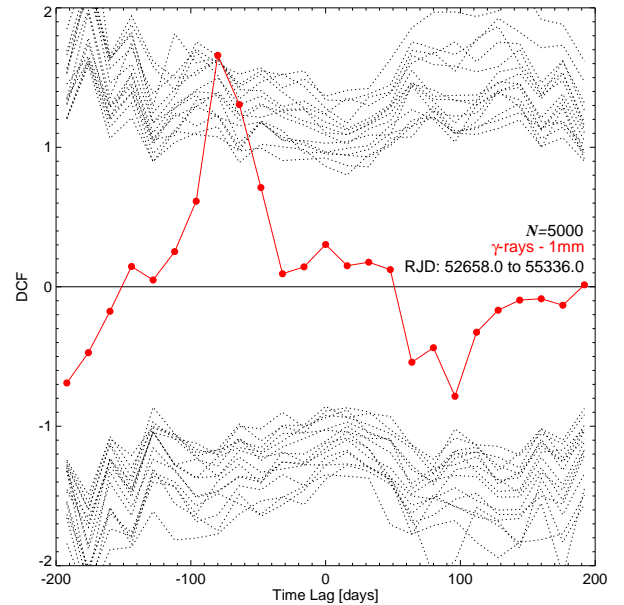}
   \caption{Discrete cross-correlation function between the $\gamma$-ray and 1\,mm light curves of OJ287 (red points). 
   The dotted curves at positive (negative) DCF values denote 99.7\% confidence limits for correlation rather than stochastic variability for all combinations of $a_{\gamma}$ and $a_{\rm{1mm}}$ discussed in the text.}
   \label{dcf}
\end{figure}

We note that the correlation peak in Fig.~\ref{dcf} is dominated by events ${\rm{B}}_{\gamma}$ and ${\rm{B}}_{\rm{mm}}$.
Despite the prominence of flare ${\rm{A}}_{\rm{mm}}$, the weakness of its $\gamma$-ray counterpart (${\rm{A}}_{\gamma}$) prevents the pair of events from being apparent in the DCF at their time lag of $\sim-120$\,days.
However, there is statistical support for the hypothesis that flares ${\rm{A}}_{\gamma}$ and ${\rm{A}}_{\rm{mm}}$ are also physically related.
The probability that a $\gamma$-ray outburst (approximated as instantaneous relative to the mm-wave flares) occurs by chance during a $\sim120$-day rise time of a 1\,mm flare peaking at $>5.5$\,Jy is only 17\%.
This probability is inferred from the duration of the long-term 1\,mm SMA light curve\footnote{\tt http://sma1.sma.hawaii.edu/callist/callist.html} and the number of such flares. 
Hence, the probability that two $\gamma$-ray flares at random times occur by chance during the rising phase of two mm-wave flares is 3\%, i.e., events ${\rm{A}}_{\rm{mm}}$ and ${\rm{B}}_{\rm{mm}}$ are associated with ${\rm{A}}_{\gamma}$ and ${\rm{B}}_{\gamma}$, respectively, at 97.0\% confidence level.
This rises to 99.2\,\% if, instead of ${\rm{A}}_{\rm{mm}}$ and ${\rm{B}}_{\rm{mm}}$, the two high $p_{\rm{C1}}$ mm-wave peaks are considered.

\section{Discussion and Conclusions}

The two 0.1--200\,GeV flares in \object{OJ287} allow us to assess the correspondence between $\gamma$-ray and lower-frequency variations. We find that two kinds of events at millimeter wavelengths are related to these $\gamma$-ray outbursts at high significance:
(1) the early, rising phases of the two most luminous 1\,mm flares ever detected in this blazar, (${\rm{A}}_{\rm{mm}}$ and ${\rm{B}}_{\rm{mm}}$); and
(2) two sharp increases to unprecedented levels of linear polarization ($\sim14$\% and $\sim22$\%) in bright jet feature C1 $>14$\,pc from the central engine.
These events also coincide with:
(3) two sharp optical flares;
(4) two peaks in optical polarization of $\sim35$\%; and
(5) the similarity of optical and mm-wave polarization position angle both during and between the flares at $\chi\approx160^{\circ}$-$170^{\circ}$.

The exceptionally high polarization of C1 during $\gamma$-ray flare ${\rm{B}}_{\gamma}$ provides extremely strong evidence that the event occurred in C1. 
This has two important implications.
First, given the distance of C1 from the central engine, the $0.1$--$200$\,GeV flares must be produced by inverse Compton (IC) scattering rather than nuclear collisions.
Second, the $\gamma$-ray IC emission arises from either the synchrotron self-Compton (SSC) process or IC scattering of infrared radiation from a hot, dusty torus of size $\sim 10$~pc \citep[IC/dust;][]{2004ApJ_600L_27B,Sokolov:2005p12701}. 
An SSC model is possible given the low ratio of $\gamma$-ray to synchrotron luminosity between $10^{14}$ and $10^{15}$\,Hz ($\approx 2$) in \object{OJ287}, based on the fluxes we measure and a spectral index of $-1.5$ between $10^{14}$ and $10^{15}$\,Hz \citep{Villforth:2010p11557}. 
Inverse-Compton scattering of dust emission predicts that the optical and $\gamma$-ray emission should vary together, since the electron energies involved are similar \citep[see, e.g.,][]{Marscher:2010p11374}.
This is the case during the $\gamma$-ray flares. 
The general data are therefore consistent with both the SSC and the IC/dust models.

\begin{figure}
   \centering
   \includegraphics[clip,width=8.cm]{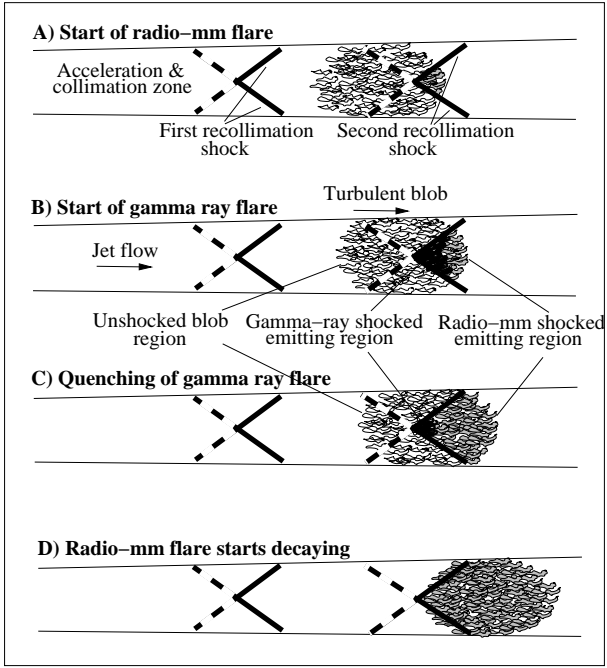}
   \caption{Sketch of the proposed model for the multi-wavelength flaring behavior of OJ 287.}
   \label{sketch}
\end{figure}

The observed behavior agrees with a scenario in which the optical and $\gamma$-ray flares are produced in C1 by particle acceleration in a moving blob when it crosses a standing shock (Fig.~\ref{sketch}) well beyond the ACZ. 
\citet{2007ApJ_659L_107D} have associated the innermost mm-wave jet emission feature with a recollimation shock at the end of the ACZ. We identify C0 as such a feature, and C1 as a second re-collimation shock, as seen in hydrodynamical simulations \citep{1997ApJ_482L_33G,Agudo:2001p460} and often in VLBA images \citep{Jorstad:2001p5655,Jorstad:2005p264}.
\citet{Jorstad:2010p11830} proposed a similar system of three conical shocks to explain the fine structure in the innermost 7\,mm jet of \object{3C~454.3} in a feature that would be identified as the ``core'' at centimeter wavelengths.
In \object{OJ287}, the higher flux of C1 relative to C0 could be caused by curvature in the jet increasing the Doppler beaming with distance from C0 so that it is maximized at C1.

A blob is either a shock or another disturbance that propagates down the jet with significantly higher relativistic electron density and magnetic field than in the ambient flow. 
The typical weakness of the observed polarization outside the high polarization peaks favors a turbulent plasma over a shock. 
\citet{Cawthorne:2006p409} shows that, in this case, the side of the conical shock nearest to the line of sight can be highly polarized with $\chi$ parallel to the jet axis. 
The remainder of the conical shock, farther from the line of sight, has much lower polarization. 
Because of light-travel delays, we first see the blob penetrate the near side, and therefore observe a major increase in polarization. 
As the outburst develops, more of the emission comes from the low-polarization far side, which decreases $p$ while the mm-wave flux density continues to increase. 
This is the pattern observed during both flares.

The mm-wave and optical flares start at essentially the same time as the magnetic field and electron energies near the leading edge of the blob become amplified as they pass the standing shock front. 
The mm-wave flux outburst continues as the relatively low-energy electrons fill the shocked region. 
The optical and $\gamma$-ray emission, produced by higher-energy electrons that cannot travel far before suffering radiative energy losses, is confined closer to the shock front where particles are accelerated \citep{1985ApJ_298_114M}. If synchrotron losses were the sole factor, the optical and $\gamma$-ray flux would reach a plateau once the shocked blob plasma fills the entire layer behind the shock, declining only when the upstream side of the blob passes the shock front. However, if SSC emission dominates, the radiative losses will increase as synchrotron photons from the flare reach electrons that scatter them to high energies. 
The radiative energy losses then increase until the loss rate reaches a factor of $[1+(L_{\rm SSC}/L_{\rm synch})]$ ($\approx 3$ in the case of \object{OJ287}) higher than the initial case of only synchrotron losses. 
This decreases the volume of emission as a function of time in the optical and $\gamma$-ray ranges. 
The time-scale for this decrease is the light-travel time across the shock times $(1+z)\delta^{-1} \sim 20 (\delta/19)^{-1} (a/0.054\,{\rm mas})$ days, where $a$ is the cross-sectional angular size of the jet and the Doppler factor ($\delta$) is that derived by \citet{Jorstad:2005p264}. 
This is similar to the time-scale of the $\gamma$-ray flux decline of flare ${\rm{B}}_{\gamma}$ ($\sim3$-4 weeks, see Fig. \ref{tflux}).

Quenching of the flares by increasing SSC energy losses operates only if the $\gamma$-rays are produced by the SSC process. 
Hence, the multi-frequency behavior of the flares is difficult to reproduce in the IC/dust model. 
Also, infrared emission from the dusty torus has not been detected thus far in BL~Lacs such as \object{OJ287}, as far as the authors know.
We thus favor the SSC mechanism.

\begin{acknowledgements}
The authors thank the referee for constructive comments.
This research was funded by NASA grants NNX08AJ64G, NNX08AU02G, NNX08AV61G, and NNX08AV65G, NSF grant AST-0907893, and NRAO award GSSP07-0009 (Boston University); RFBR grant 09-02-00092 (St.~Petersburg State University); MICIIN grants AYA2007-67627-C03-03 and AYA2010-14844, and CEIC (Andaluc\'{i}a) grant P09-FQM-4784 (IAA-CSIC); the Academy of Finland
(Mets\"{a}hovi); and NASA grants NNX08AW56S and NNX09AU10G (Steward Observatory).
The VLBA is an instrument of the NRAO, a facility of the NSF operated under cooperative agreement by AUI. 
The PRISM camera at Lowell Observatory was developed by Janes et al., with funding from the NSF, Boston University, and Lowell Observatory. The Calar Alto Observatory is jointly operated by MPIA and IAA-CSIC. 
The IRAM 30\,m Telescope is supported by INSU/CNRS (France), MPG (Germany), and IGN (Spain).
The Submillimeter Array is a joint project between the SAO and the Academia Sinica. 
\end{acknowledgements}



\begin{thebibliography}{}

\bibitem[\protect\citeauthoryear{Abdo et~al.}{Abdo
  et~al.}{2010a}]{Abdo:2010p11947}
Abdo, A.~A., et~al. 2010a, \apj, 716, 30

\bibitem[\protect\citeauthoryear{Abdo et~al.}{Abdo
  et~al.}{2010b}]{Abdo:2010p12082}
Abdo, A.~A., et~al. 2010b, \apjs, 188, 405

\bibitem[\protect\citeauthoryear{Abdo et~al.}{Abdo
  et~al.}{2010c}]{Abdo:2010p11698}
Abdo, A.~A.,et~al. 2010c, \apj, 772, 520

\bibitem[\protect\citeauthoryear{Abdo et~al.}{Abdo
  et~al.}{2010d}]{Abdo:2010p11811}
Abdo, A.~A., et~al. 2010d, \nat, 463, 919

\bibitem[\protect\citeauthoryear{Agudo et~al.}{Agudo
  et~al.}{2001}]{Agudo:2001p460}
Agudo, I., et~al. 2001, \apj, 549, L183

\bibitem[\protect\citeauthoryear{Agudo et~al.}{Agudo
  et~al.}{2006}]{Agudo:2006p203}
Agudo, I., et~al. 2006, \aap, 456, 117

\bibitem[\protect\citeauthoryear{Agudo et~al.}{Agudo
  et~al.}{2007}]{Agudo:2007p132}
Agudo, I., et~al. 2007, \aap, 476, L17

\bibitem[\protect\citeauthoryear{Agudo et~al.}{Agudo
  et~al.}{2010}]{Agudo:2010p12104}
Agudo, I., Thum, C., Wiesemeyer, H.,  \& Krichbaum, T.~P.  2010, \apjs, 189, 1

\bibitem[\protect\citeauthoryear{B{\l}a{\.z}ejowski et~al.}{B{\l}a{\.z}ejowski
  et~al.}{2004}]{2004ApJ_600L_27B}
B{\l}a{\.z}ejowski, M., Siemiginowska, A., Sikora, M., Moderski, R.,  \&
  Bechtold, J. 2004, \apj, 600, L27

\bibitem[\protect\citeauthoryear{Cawthorne}{Cawthorne}{2006}]
{Cawthorne:2006p409}
Cawthorne, T.~V. 2006, \mnras, 367,  851

\bibitem[\protect\citeauthoryear{Chatterjee et~al.}{Chatterjee
  et~al.}{2008}]{2008ApJ_689_79C}
Chatterjee, R., et~al. 2008, \apj, 689, 79

\bibitem[\protect\citeauthoryear{D'Arcangelo et~al.}{D'Arcangelo
  et~al.}{2007}]{2007ApJ_659L_107D}
D'Arcangelo, F.~D., et~al. 2007, \apj, 659, L107

\bibitem[\protect\citeauthoryear{Edelson \& Krolik}{Edelson \&
  Krolik}{1988}]{Edelson:1988p12129}
Edelson, R.~A.,  \& Krolik, J.~H. 1988, \apj, 333, 646

\bibitem[\protect\citeauthoryear{Foschini et~al.}{Foschini
  et~al.}{2010}]{Foschini:2010p12452}
Foschini, L., et~al. 2010, \mnras, 408, 448

\bibitem[\protect\citeauthoryear{G\'{o}mez et~al.}{G\'{o}mez et~al.}
{1997}]{1997ApJ_482L_33G}
G\'{o}mez et~al. 1997, \apj, 482, L33

\bibitem[\protect\citeauthoryear{Gurwell et~al.}{Gurwell
  et~al.}{2007}]{Gurwell:2007p12057}
Gurwell, M.~A., Peck, A.~B., Hostler, S.~R., Darrah, M.~R.,  \& Katz, C.~A.
  2007, in ASP Conf. Ser. 375, From Z-Machines to ALMA: (Sub)millimeter 
  Spectroscopy of Galaxies, ed. A. J. Baker et al. (San Francisco, CA: ASP), 
  234
  
\bibitem[\protect\citeauthoryear{Hufnagel \& Bregman}{Hufnagel \& 
Bregman}{1992}]{Hufnagel:1992p13974}
Hufnagel, B.~R., \& Bregman, J.~N., 1992, \apj, 386, 473

\bibitem[\protect\citeauthoryear{Jones}{Jones}{1988}]{Jones:1988p7083}
Jones, T.~W. 1988, \apj, 332, 678

\bibitem[\protect\citeauthoryear{Jorstad et~al.}{Jorstad
  et~al.}{2001a}]{Jorstad:2001p5655}
Jorstad, S.~G.,  et~al. 2001a, \apjs, 134, 181

\bibitem[\protect\citeauthoryear{Jorstad et~al.}{Jorstad
  et~al.}{2001b}]{Jorstad:2001b}
Jorstad, S.~G.,  et~al. 2001b, \apj, 556, 738

\bibitem[\protect\citeauthoryear{Jorstad et~al.}{Jorstad
  et~al.}{2005}]{Jorstad:2005p264}
Jorstad, S.~G., et~al. 2005, \aj, 130, 1418

\bibitem[\protect\citeauthoryear{Jorstad et~al.}{Jorstad
  et~al.}{2007}]{2007AJ.134.799J}
Jorstad, S.~G., et~al. 2007, \aj, 134, 799

\bibitem[\protect\citeauthoryear{Jorstad et~al.}{Jorstad
  et~al.}{2010}]{Jorstad:2010p11830}
Jorstad, S.~G., et~al. 2010, \apj, 715, 362

\bibitem[\protect\citeauthoryear{L{\"a}hteenm{\"a}ki \&
  Valtaoja}{L{\"a}hteenm{\"a}ki \& Valtaoja}{2003}]{Lahteenmaki:2003p5657}
L{\"a}hteenm{\"a}ki, A.,  \& Valtaoja, E. 2003, \apj, 590, 95

\bibitem[\protect\citeauthoryear{Marscher \& Gear}{Marscher \&
  Gear}{1985}]{1985ApJ_298_114M}
Marscher, A.~P.,  \& Gear, W.~K. 1985, \apj, 298, 114

\bibitem[\protect\citeauthoryear{Marscher et~al.}{Marscher
  et~al.}{2008}]{2008Natur_452_966M}
Marscher, A.~P., et~al. 2008, \nat, 452, 966

\bibitem[\protect\citeauthoryear{Marscher \& Jorstad}{Marscher \&
  Jorstad}{2010}]{Marscher:2010p12402}
Marscher, A.~P.,  \& Jorstad, S.~G. 2010, in Fermi meets Jansky -- 
 AGN at Radio and Gamma-Rays, ed. T. Savolainen et al., 171
  
\bibitem[\protect\citeauthoryear{Marscher et~al.}{Marscher
  et~al.}{2010}]{Marscher:2010p11374}
Marscher, A.~P., et~al. 2010, \apjl, 710, L126

\bibitem[\protect\citeauthoryear{Mattox et~al.}{Mattox
  et~al.}{1997}]{Mattox:1997p12917}
Mattox, J.~R.,  et~al. 1997, \apj, 476, 692

\bibitem[\protect\citeauthoryear{Max-Moerbeck et~al.}{Max-Moerbeck
  et~al.}{2010}]{MaxMoerbeck:2010p12574}
Max-Moerbeck, W., et~al. 2010, in Fermi meets Jansky -- 
 AGN at Radio and Gamma-Rays, ed. T. Savolainen et al., 77
  
\bibitem[\protect\citeauthoryear{Pushkarev et~al.}{Pushkarev 
et~al.}{2009}]{Pushkarev:2009p9412}
 Pushkarev, A.~B., Kovalev, Y.~Y., Lister, M.~L. \& Savolainen, T.  
 2009, \aap, 507, L33

\bibitem[\protect\citeauthoryear{Sokolov \& Marscher}{Sokolov \&
  Marscher}{2005}]{Sokolov:2005p12701}
Sokolov, A.,  \& Marscher, A.~P. 2005, \apj, 629, 52

\bibitem[\protect\citeauthoryear{Tateyama \& Kingham}{Tateyama \&
  Kingham}{2004}]{Tateyama:2004p11950}
Tateyama, C.~E.,  \& Kingham, K.~A. 2004, \apj, 608, 149

\bibitem[\protect\citeauthoryear{Ter\"{a}sranta et~al.}{Ter\"{a}sranta
  et~al.}{1998}]{1998A&AS_132_305T}
Ter\"{a}sranta, H., et~al. 1998, \aaps, 132, 305

\bibitem[\protect\citeauthoryear{Timmer \& Koenig}{Timmer \&
  Koenig}{1995}]{Timmer:1995p12152}
Timmer, J.,  \& Koenig, M. 1995, \aap, 300, 707

\bibitem[\protect\citeauthoryear{Villforth et~al.}{Villforth
  et~al.}{2010}]{Villforth:2010p11557}
Villforth, C., et~al. 2010, \mnras, 402, 2087

\end{thebibliography}
\end{document}